\documentclass[twocolumn,aps,prb,groupedaddress,showpacs]{revtex4}
\usepackage[T1]{fontenc}
\usepackage{graphicx}% Include figure files

\def\pes{Pb$_{1-x}$Eu$_x$Se}

\begin{document}
\title{MAGNETIC CONTRIBUTION TO THE SPECIFIC HEAT OF
Pb$_{1-x}$Eu$_x$Te}

\author{M.~G\'orska}
\affiliation{Institute of Physics, Polish Academy of Sciences,
 Al. Lotnik\'{o}w 32/46, 02-668 Warsaw, Poland}
\author{A.~{\L}usakowski}
\affiliation{Institute of Physics, Polish Academy of Sciences,
 Al. Lotnik\'{o}w 32/46, 02-668 Warsaw, Poland}
\author{A.~J\c{e}drzejczak}
\affiliation{Institute of Physics, Polish Academy of Sciences,
 Al. Lotnik\'{o}w 32/46, 02-668 Warsaw, Poland}
\author{Z.~Go{\l}acki}
\affiliation{Institute of Physics, Polish Academy of Sciences,
Al. Lotnik\'{o}w 32/46, 02-668 Warsaw, Poland}
\author{J.R. Anderson}
\affiliation{Department of Physics, University of Maryland,
College Park, MD 20742}
\author{H. Balci}
\altaffiliation[Present address: ] {Department of Physics,
University of Illinois, Urbana, Illinois 61801}
\affiliation{Department of Physics, University of Maryland,
College Park, MD 20742}
\author{R.R.~Ga{\l}\c{a}zka}
\affiliation{Institute of Physics, Polish Academy of Sciences,
        Al. Lotnik\'{o}w 32/46, 02-668 Warsaw, Poland}
\begin{abstract}
The temperature dependence of the magnetic specific heat has been
studied experimentally and theoretically in the semimagnetic  
semiconductor {Pb$_{1-x}$Eu$_x$Te} for  $x$=0.027 and $x$=0.073, over the  
temperature range from 0.5 K to 10 K, in magnetic fields up to 2 T.
There was a maximum in the magnetic specific heat between 1
and 3 K even in zero and low magnetic fields; this maximum shifted toward
higher temperatures with increasing  magnetic field. 
The experimental data have been analyzed in the framework of a model
in which we assume that the ground states of europium ions are split
even without an external magnetic field. We present arguments which
support this assumption and we show that it is possible to find a
physical mechanism leading to the splitting which can explain the
experimental results. 

\end{abstract}

%\keywords{}

\maketitle

\section{Introduction}
Semimagnetic semiconductors (SMSs), also known as diluted magnetic
semiconductors (DMS), have been studied extensively during
the past two decades. Recently, there has been a considerable interest
in these materials because of their possible applications
in spintronics. Optical and magnetic measurements have
shown that, in general, IV-VI SMSs with a 3$d$ element as
the magnetic ion have a much weaker exchange interaction
than that found in II-VI SMSs with the same magnetic ions.
Also, the IV-VI SMSs with rare earths have a weaker
exchange interaction than that found in the same materials
with a 3$d$ element as the magnetic ion (for a review
see Refs. \onlinecite{wd03,ts03,jf88,dj95}).
Electron spin resonance investigations in SnTe with Mn,
Eu, and Gd have shown that the exchange interaction between free carriers
and magnetic ions is roughly an order of magnitude smaller for Eu and Gd
than for Mn. \cite{pu75}
The mechanism of the exchange interaction among magnetic ions
in IV-VI SMSs is still not well understood.

     Our previous investigations of the magnetic properties of Pb$_{1-x}$Mn$_x$Te \ 
and {Pb$_{1-x}$Eu$_x$Te} indicated a small ($J/k_B < $ 1 K), antiferromagnetic
exchange interaction among magnetic ions.\cite{mg88,mg90,mg97}
In {Pb$_{1-x}$Eu$_x$Te} the absolute value of the exchange constant was about three
times smaller than in Pb$_{1-x}$Mn$_x$Te \   and decreased with the increasing
Eu-content. In Ref. \onlinecite{th97} ter Haar {\it et al.}
have observed magnetization steps in the high-field
magnetization at milikelvin temperatures in {Pb$_{1-x}$Eu$_x$Te} and found
exchange constant values similar to ours.
By comparison with the results in II-VI SMSs we came to a conclusion,
that in IV-VI SMSs the dominant exchange mechanism is the
superexchange between nearest neighbors (NN). In order to
develop a more complete model and to obtain parameters
for the exchange interaction, we have made complementary measurements
of the magnetic specific heat of Pb$_{1-x}$Mn$_x$Te \   and analyzed the results
together with the results of the magnetization and magnetic susceptibility
measurements. \cite{al02}  It turned out that the mechanism of
the exchange interaction in Pb$_{1-x}$Mn$_x$Te \   may be more complex than just
the NN superexchange. To explain the temperature and magnetic
field dependence of the specific heat of Pb$_{1-x}$Mn$_x$Te \   it was necessary
to take into account a splitting of the ground-energy state
of single Mn-ions in Pb$_{1-x}$Mn$_x$Te \  and the $p-d$ coupling between
magnetic ion spins and free carriers.

In the present paper we report studies of the
magnetic specific heat of {Pb$_{1-x}$Eu$_x$Te} crystals and compare the results
with those obtained in Pb$_{1-x}$Mn$_x$Te.  Some preliminary data
have been recently reported. \cite{mg04}  This is a first
investigation of the magnetic contribution to the specific
heat in rare-earth-doped SMSs.  In the following sections
we present the experimental results and analysis of 
specific heat measurements and give a physical model for 
the magnetic specific heat of {Pb$_{1-x}$Eu$_x$Te}. 

In theoretical analysis presented in Section III we argue that the
experimentally observed 
magnetic specific heat is mainly due to single europium ions split in
the disordered crystal environment. The splittings of magnetic ions
caused by the crystal field are  
small and have been usually detected in EPR experiments. However, for
europium in IV-VI semimagnetic semiconductors these splittings are large enough to
be observed also in different kinds of measurements. In closely related
semiconductors Pb$_{1-x}$Eu$_x$S and \pes\  magnetization steps 
due to the splitting of single europium ions have been
observed.\cite{bindilatti1,bindilatti2} In Pb$_{1-x}$Eu$_x$Te we expect a similar effect.

There is no consensus in the literature concerning the mechanism of 
the ground state splitting of $^8S$ rare earth ions in crystals. Several models
were proposed and analyzed.\cite{wybourne,barnes,lusakowski1} It seems
that at present it is impossible to decide unequivocally which mechanism 
should be applied to a specific case. Instead, we suspect that in
each case several mechanisms should be considered. 
Therefore, in Section III we consider four different physical mechanisms leading
to the splitting and estimate  magnitudes of the resulting 
splittings.

The important part of the theoretical analysis is the incorporation of
 deformations 
of the Pb$_{1-x}$Eu$_x$Te crystal.  These deformations are caused by the difference
between Eu and 
Pb atoms and it turns out that even relatively small departures from
octahedral symmetry may lead to splittings of the order 1 -
10~K. Therefore, they should be taken into account in analysis of magnetic
specific heat measurements.

The theoretical considerations in the present paper are limited mainly to the case
of the $x$=0.027 sample. For this sample most of the magnetic ions are
singles, i.e. they have no nearest magnetic neighbors. For the
sample with higher concentration of Eu, $x$=0.073, 
although the theory may be applied formally, its quantitative
predictions are not very reliable, because for such a high concentration a 
significant number of the Eu atoms is in larger, many atom clusters and the
theoretical analysis of such a system is much more difficult. 

\section{Experiment}
  We have measured the specific heat of {Pb$_{1-x}$Eu$_x$Te} with $x$ values of
0.027 and 0.073.
The samples of {Pb$_{1-x}$Eu$_x$Te} were grown by the Bridgman technique and the Eu concentration
was estimated from the amounts of the components introduced into
the growth chamber and measured by energy dispersive x-ray analysis (EDAX).
The nominal $x$-values were 0.03 and 0.06 with uncertainty
of about 20\%. The crystals were cut
in the shape of Hall bars with typical dimensions 1.5 $\times$ 2 $\times$
6 mm$^3$. The samples were $p$ type with carrier concentrations,
from Hall measurements, of about 1 $\times$ 10$^{18}$ cm$^{-3}$.
With increasing $x$ the hole concentration decreased and the mobility
increased.

     Previously we have measured high-temperature magnetic susceptibility
and low-temperature, high-field magnetization of {Pb$_{1-x}$Eu$_x$Te} with
$x$ up to 0.1. \cite{mg90,mg97} By fitting the susceptibility data
to the Curie-Weiss law we obtained
the average Eu content in our samples $(x_{av})$ and very small Curie-Weiss
temperatures indicating an antiferromagnetic exchange between
Eu ions, $J/k_B$ = -0.38~K and -0.27~K for $x_{av}$ = 0.027 and 0.073,
respectively.

     The measurements of the heat capacity were performed in a cryostat
using  $^3$He and $^4$He systems over the temperature range 0.5 - 15 K
in magnetic fields 0, 0.5, 1, and 2 T. We used the standard
adiabatic heat-pulse method. \cite{fm63} Errors in the heat capacity
values were about 5\%. The experimental details
have been described elsewhere. \cite{al02}

     In order to obtain the magnetic contribution to the specific
heat, $C_H$, it was necessary to subtract the specific heat of the
{Pb$_{1-x}$Eu$_x$Te} lattice from the measured total specific heat of {Pb$_{1-x}$Eu$_x$Te}.
This was not a simple task. Bevolo {\it et al.} found that the
specific heat of PbTe has an anomaly below 5 K and could not be
fitted with the standard expression $C = \gamma T + \alpha T^3$,
where $\gamma T$ and $\alpha T^3$ are the electronic and lattice
contributions, respectively.\cite{bs76} In fact, they could not
obtain a satisfactory fit to their data with an expression of the
form, $C = \gamma T + \alpha T^3 + \sum_{i=1}^n
\delta_{i}T^{2i+3}$ unless $n$ was at least 10. Therefore, we
measured the heat capacity of our own Bridgman-grown PbTe sample
in zero magnetic field and 2 T, over the temperature range from
0.5 to 15 K and found that the temperature dependence was the same
for 0 and 2 T within our experimental error (as expected). In Ref.
\onlinecite{al02} we described this experiment and have shown the
specific heat of PbTe vs temperature in zero magnetic field. In
our preliminary paper we have shown the result of simple
subtraction of the PbTe specific heat from the {Pb$_{1-x}$Eu$_x$Te} specific
heat. \cite{mg04} In the present paper we take into account the
effect that the replacement of Pb with an atomic mass of 207.2 by
Eu with an atomic mass of 151.97 leads to a decrease in heat
capacity, even for small values of $x$. To account for this we
divided the entire set of PbTe specific heat data by empirically
determined factors, 1.11 for $x$=0.073 and 1.04 for $x$=0.027,
before subtracting from the {Pb$_{1-x}$Eu$_x$Te}. These factors were determined
by assuming that at temperatures above 15 K, in the absence of an
applied magnetic field, the magnetic contribution to the specific
heat of {Pb$_{1-x}$Eu$_x$Te} is negligible. Therefore, this division by 1.11
(1.04) gave results for PbTe that were the same as those for
{Pb$_{1-x}$Eu$_x$Te} at 15 K for $x$ = 0.073 (0.027). Since this is an empirical
correction, we emphasize in the present work the data at
temperatures below 5 K where the lattice specific heat is much
smaller than the total specific heat. In the interesting region,
below 2 K, the specific heat of PbTe was more than 3 orders of
magnitude smaller than that of Pb$_{1-x}$Eu$_x$Te.

\begin{figure}
\includegraphics[scale=0.3,angle=0]{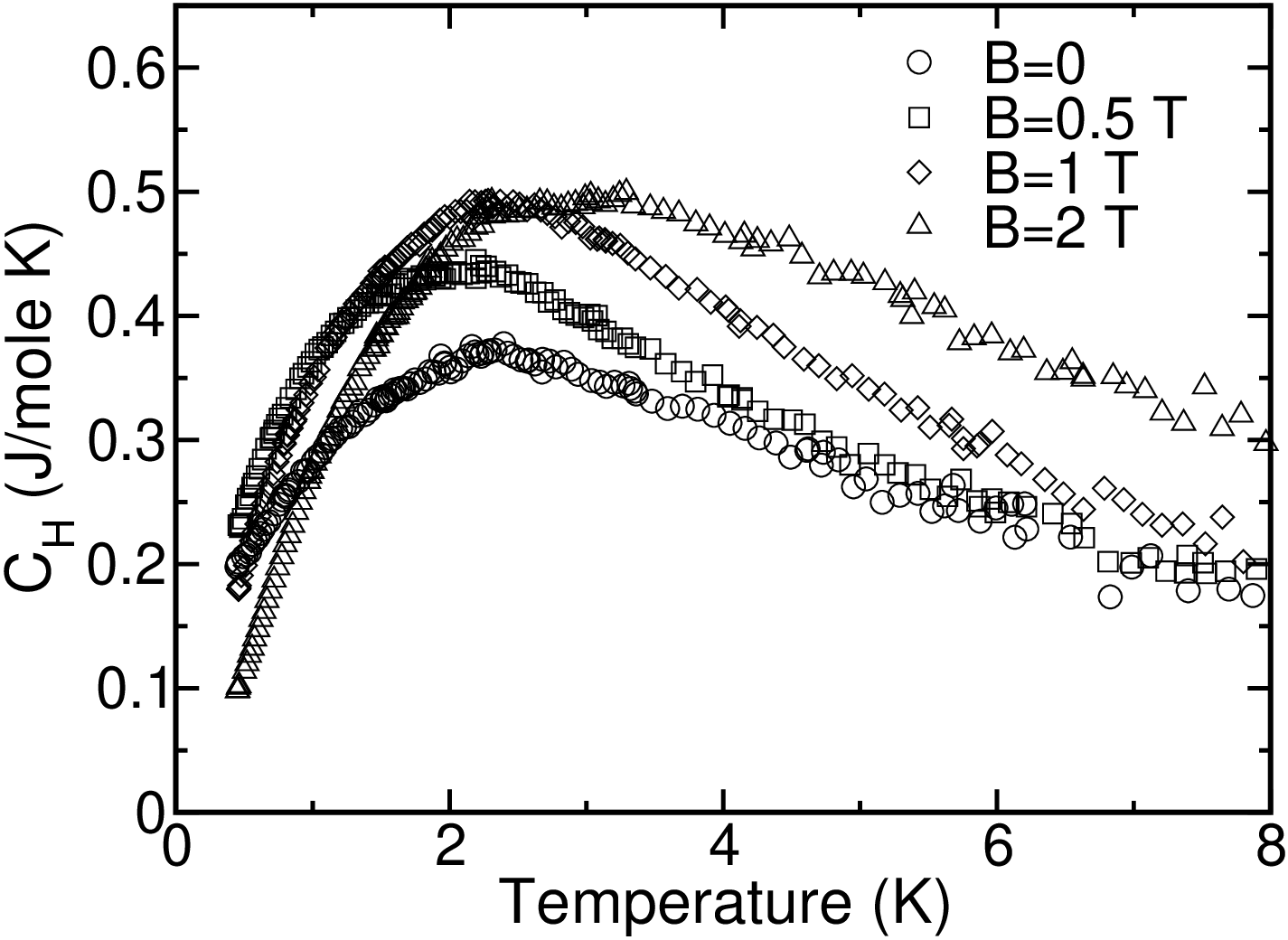}
\caption{\label{exp1} Magnetic specific heat of {Pb$_{1-x}$Eu$_x$Te} with x = 0.073 in various
magnetic fields. }
\label{fig1}
\end{figure}

\begin{figure}
\vspace{0.5cm}
\begin{center}
\includegraphics*[scale=0.3,angle=0]{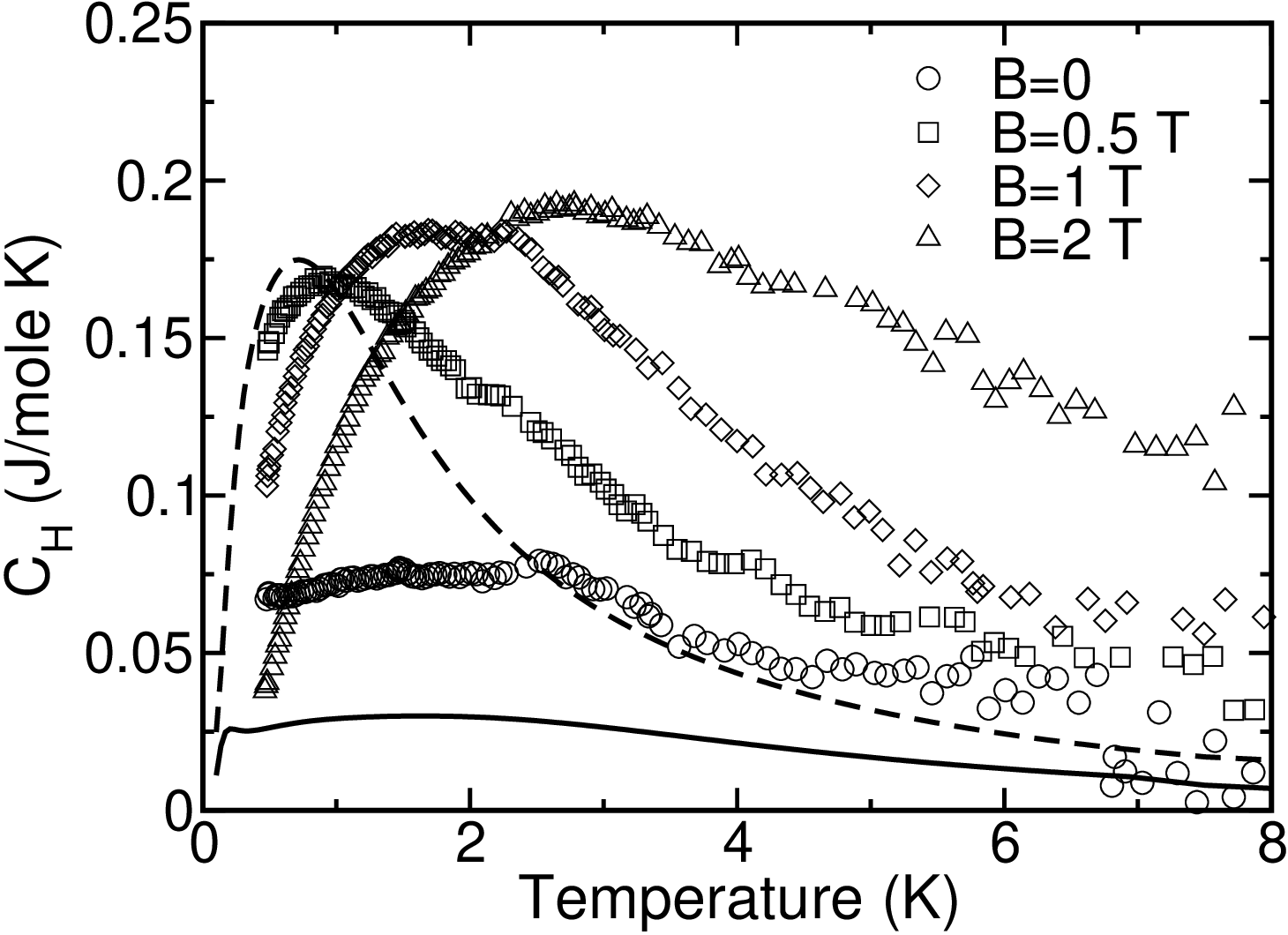}% Here is how to import EPS art
\end{center}
\caption{\label{ft1} Magnetic specific heat of {Pb$_{1-x}$Eu$_x$Te} with x = 0.027 in various
magnetic fields. Points - experimental data, lines -
theoretical predictions of the nearest neighbours interaction model 
for B=0 (continuous line) and B=0.5~T (broken line). }
\end{figure}

The magnetic specific heat data for {Pb$_{1-x}$Eu$_x$Te} are shown in Figs.\
\ref{fig1} and \ref{ft1}. We believe that the scatter in the data
represents the experimental error. 
For both $x$-values there is a broad maximum in the magnetic
specific heat at about 2 K in zero magnetic field. The maximum
is several times higher than that predicted by the cluster model of
superexchange interaction between nearest neighbors.
At higher magnetic fields the value of specific heat at the maximum
increases and above 0.5 T it shifts to higher temperatures;
for $x$ = 0.073 the shift is smaller than for $x$ = 0.027. This
behavior is different from that observed in Pb$_{1-x}$Mn$_x$Te \cite{al02}
or  Sn$_{1-x}$Mn$_x$Te, \cite{pe94} where the
value of the magnetic specific heat at zero magnetic field
and all higher fields was nearly the same, but the shift
of the maximum with increasing magnetic field was bigger
than in {Pb$_{1-x}$Eu$_x$Te}. 

\section{Theoretical analysis}
Before we start the analysis of magnetic specific heat (MSH) 
let us recall the model of the Europium atom in Pb$_{1-x}$Eu$_x$Te which describes quite
well magnetization and magnetic susceptibility experimental
data.\cite{mg97}

The electron configuration of a free Eu atom is $4f^75s^25p^66s^2$. It is
believed that when it replaces a Pb atom in PbTe the electrons from
the outermost shell, $6s^2$, play the role of  $6p^2$ electrons of Pb and
contribute to crystal bindings. Due to the strong relativistic
downward shift, the energy position of $6s^2$ Pb
electrons is deeply in the valence band.\cite{dietl1} As a result we
obtain Eu$^{2+}$ ion, 
electrically inactive with respect to the crystal, despite the fact
that it replaces Pb atom from IV group. Such a picture is
confirmed  by the fact that the presence of even 10\% of Eu atoms has
almost no effect on carrier concentration.\cite{ts03} \\
According to Hund's rule the ground state of Eu ion is $^8S$, it means
that seven $4f$ electrons form the spin S=7/2 and the total angular
momentum L=0. Thus, the ground state of the ion is eightfold
degenerate. This degeneracy is removed by external magnetic
field, {\bf B} or by interaction with another magnetic ion. With such
assumptions the Hamiltonian for the spin subsystem reads
\begin{equation}
\label{t1}
H=g\mu_B\sum_{i}{\bf B}\cdot {\bf S}_i-\sum_{ij}J_{ij}{\bf S}_i\cdot {\bf S}_j,
\end{equation}
where the $g$-factor $g=2$, $\mu_B$ is the Bohr magneton, and $J_{ij}$ is
the exchange integral between the $i$-th and $j$-th spins.  
If the content of
Eu is small, of the order of 1-3\%, we may safely assume that most of
the ions have no nearest magnetic neighbors and only small
percentage of them form two or three-atom clusters called, in the literature,
pairs, open triangles, and closed triangles. Assuming the statistical
distribution of Eu atoms in the PbTe lattice, 
one knows the average number of singles and the average number of atoms in pairs, open
triangles, and closed triangles.\cite{behringer} Then every
thermodynamic quantity, in particular the magnetic specific heat, may be calculated.

Although such a cluster model is successful in description of
magnetization in Pb$_{1-x}$Eu$_x$Te\cite{mg97} it fails in the case of
MSH. In Fig.~\ref{ft1} we see that the calculated MSH 
is much smaller than the one observed experimentally. The
calculated MSH in Fig.~\ref{ft1}, for $B$=0 is due to pairs and
triples only, because the contribution from singles, which are for $B=0$
eight fold degenerate, is zero. The specific heat 
due to larger clusters has, for a sample with $x$=0.027, a very small
contribution of about 2.5 \%. We also think that theories like
extended nearest neighbor pair approximation (ENNPA)\cite{twardowski}
based on the long range mechanism of spin - spin interaction, are not
applicable to Pb$_{1-x}$Eu$_x$Te because 
even the nearest neighbor Eu-Eu exchange integral in Pb$_{1-x}$Eu$_x$Te is 
small, $J/k_B\approx$-0.25 K\cite{mg97,th97}, and more distant interactions, which
quickly decay with the distance, cannot explain the broad maximum for
MSH for B=0. Let us also note, that for nonzero magnetic field
the theoretical curve above 1~K lies well below the
experimental points. This 
picture gives evidence for a non-negligible  density of
energy states in the energy region far above 2~K, which must be taken
into account to describe the experiment properly. This density of
states of the 
system does not result from the model described by the Hamiltonian
Eq. (\ref{t1}).

\begin{figure}
\vspace{0.5cm}
\begin{center}
\includegraphics*[scale=0.3,angle=0]{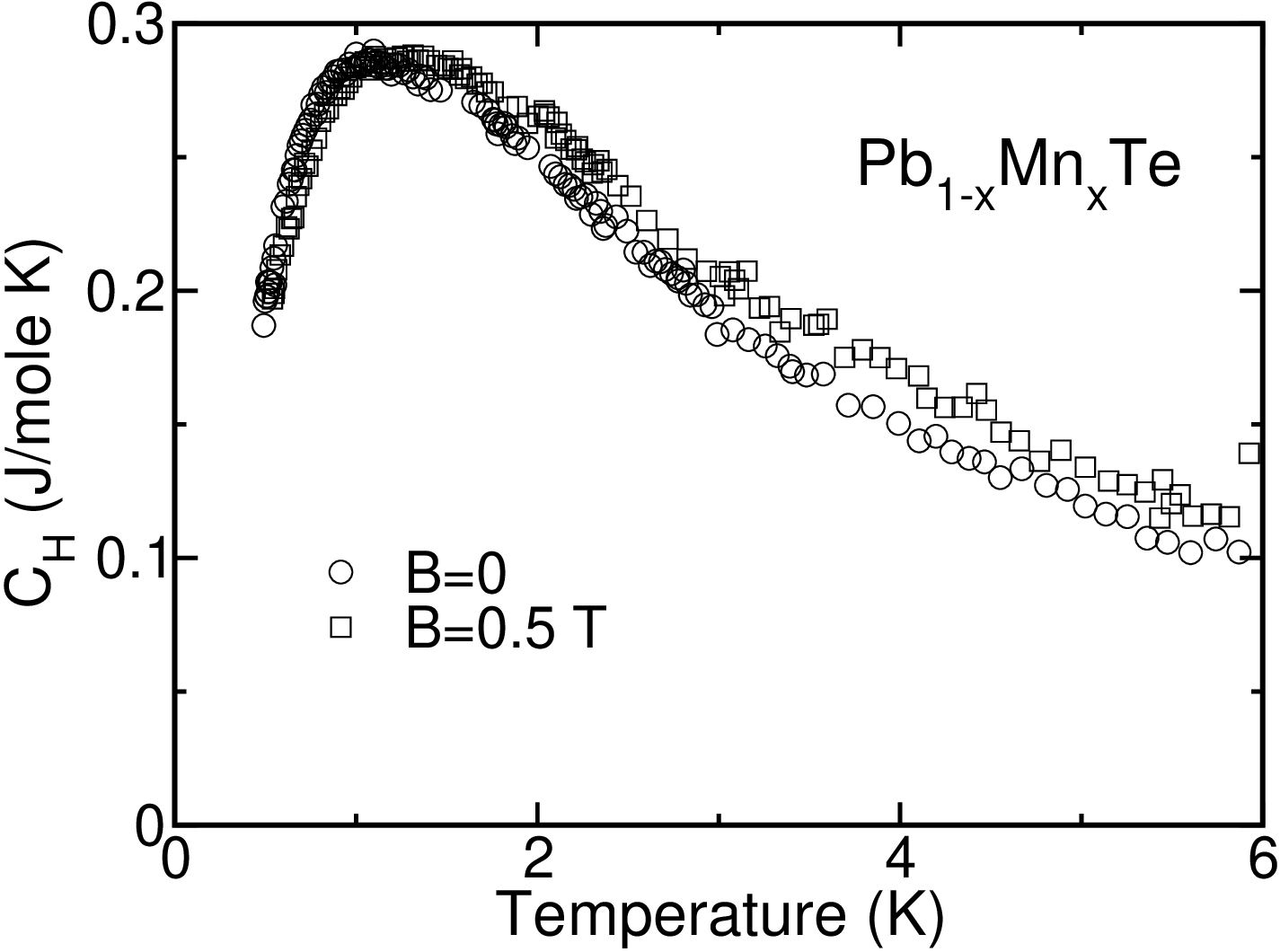}% Here is how to import EPS art
\end{center}
\caption{\label{ft2} Magnetic specific heat in Pb$_{1-x}$Mn$_x$Te\  with x=0.056 in
two magnetic fields (Ref. \onlinecite{al02}). }
\end{figure}
The situation is somewhat similar to the case of Pb$_{1-x}$Mn$_x$Te, 
where the peak in zero magnetic field MSH is also unexpectedly
high.\cite{al02} There is, however, at least one important
difference between Pb$_{1-x}$Eu$_x$Te and Pb$_{1-x}$Mn$_x$Te. According
to statistical mechanics, for any physical 
model describing MSH of a spin system, in particular for a model
described by Eq.~(\ref{t1}), we have following entropy relation:
\begin{equation}
\label{t2}
\int_{0}^{\infty}dT\frac{C_H(T)}{T}=xR\ln(2S+1)-k_B\ln(g_H),
\end{equation}
where $R$ is the molar gas constant, $S$ is the magnetic ion spin (for
Eu$^{2+}$ $S=7/2$), $C_H(T)$ is
the molar magnetic 
specific heat at temperature $T$ and in an external magnetic field, and $g_H$
is the degeneracy of the  ground state of the 
system. The degeneracy depends on the magnetic field. In an external
magnetic field $B\ne 0$,  $g_H$=1 and the second term on
the right hand  side disappears; thus calculating the difference
between 
the left hand side of Eq.~(\ref{t2}) for magnetic fields $B\ne 0$ and $B=0$ we obtain
information about the degeneracy of the ground state in zero external
magnetic field. Estimations, based
on experimental results presented in Ref.~\onlinecite{al02},
suggest that in the case of Pb$_{1-x}$Mn$_x$Te \  this difference is nearly zero (see
Fig.~\ref{ft2}). Thus, in the case of 
Pb$_{1-x}$Mn$_x$Te, any model leading to $g_{H=0} \ne 1$ must be rejected. This is not
the case of Pb$_{1-x}$Eu$_x$Te. In Fig.~\ref{ft1} we see that the difference 
between MSH measured in $B=0$ and in $B=0.5$~T is much bigger in Pb$_{1-x}$Eu$_x$Te
than in Pb$_{1-x}$Mn$_x$Te. Assuming, like in Ref.~\onlinecite{al02},
the linear temperature
dependence of  MSH below 0.5~K  we find that the
value of the integral for $B=0.5$~T 
is equal to 0.41 J/mole K. For $x$=0.027  
the right hand side of Eq.~(\ref{t2}) returns the value 0.47~J/mole K. The agreement
is quite good. For $B=0$, with the same assumption 
concerning the behavior of MSH at the lowest temperatures, the value
of the integral 
is 0.2 J/mole K.  Because the first term on the right hand side,
$xR\ln(2S+1)$, does not 
change, it means that the second one, $k_B\ln(g_{H})$, for $B=0$ must be
positive. Thus, the
experimental data suggest that in the case of Pb$_{1-x}$Eu$_x$Te in 
magnetic field $B$=0 we have non-negligible degeneracy of the ground
state of the spin system. This is the important difference between Pb$_{1-x}$Mn$_x$Te\  and
Pb$_{1-x}$Eu$_x$Te. In Pb$_{1-x}$Mn$_x$Te \  even in $B$=0 the ground state is
non-degenerate. The lack of degeneracy in Pb$_{1-x}$Mn$_x$Te \  has been discussed 
in \mbox{Ref. \onlinecite{al02}}.

At this point some clarifying remarks are necessary. First, we
do {\it not} claim 
that Pb$_{1-x}$Eu$_x$Te is a system contradicting the third law of
thermodynamics. In reality, if we take into account {\it all}
interactions and system degrees of freedom, the ground state is
non-degenerate. However, in our description we limit considerations to the spin
subsystem and   Eq.~(\ref{t2}) is
derived and may be applied only to such a subsystem. Secondly, we have
no data at temperatures below 0.5~K,  
a region which may significantly contribute to the value of the
integral.  That is why
the estimations of the left hand side of Eq.~\ref{t2} are only
semiquantitative and cannot serve as a rigorous justification 
of the approach introduced below. However, we think that they provide
important insight into the problem and  to some degree confirm the
considerations given below.

In the present paper we propose that the experimentally observed MSH in
zero magnetic field is due to the splitting of the energy levels of the single Eu ions. From the
discussion of different aspects of splitting that are presented later, it
turns out that this 
splitting is caused primarily by two mechanisms: the disordered
crystal field potential, which 
leads to virtual $4f^7 \rightarrow 4f^65d^1$ transitions, and the
internal spin - orbit coupling on $4f$ shell in the excited, $4f^65d^1$,
state.

According to the Kramers theorem, the
ground state of a single ion (of a system consisting of seven, i.e. 
an odd number of electrons
on $4f$ shell) is at least two-fold degenerate  in the absence of an external
magnetic field. Thus, 
the ground state of the spin subsystem, a significant part of which consists
of such split noninteracting ions, is also degenerate. Such an approach
is in accordance 
with the experimentally observed difference between the integrals,
Eq.~(\ref{t2}), for zero and nonzero magnetic fields. On the other
hand, the split singles contribute to the magnetic specific heat; therefore we
expect our model describing the experiment to be better than the
calculated curves in Fig.~\ref{ft1}. 
In the following analysis an important role is played by $5d$ levels of europium. 
Let us discuss now the origin and the meaning of these states involved in the
excited configuration $4f^65d^1$ of an ion.

In the investigations of Pb$_{1-x}$Eu$_x$Te by Krenn {\it et al.}\cite{krenn} the
optical dipole transitions from the $4f$ level of Eu to the vicinity
of the bottom of  conduction
band were studied. The photon energy of these transitions for Eu
concentrations corresponding to ours  was found  to
be less than 1 eV. Notice that this kind of transition may
take place only between states of different parity. Because the $f$ functions
have odd parity, the states near the bottom of the conduction band must
contain states of even parity. In pure PbTe the
wave functions of the bottom of the conduction band, of $L_6^-$ symmetry, are
of odd parity. Thus, to explain the existence of the optical transition we must
assume that in Pb$_{1-x}$Eu$_x$Te the wave functions in the vicinity of the bottom
of conduction band have certain components of even parity. The question
arises about the origin of these components. The most natural candidates
are $5d$ levels of Eu. Since it is known that in EuTe the conduction band is
built mainly from $5d$ and $6s$ states of europium,\cite{wachter} 
one may expect that the addition of a few percent of Eu to PbTe
will result in a contribution of $5d$ states to the conduction band
states. Of course, the presence of even parity states may be also
related to the disorder introduced by addition of Eu atoms to
PbTe because, strictly speaking, the 
group theory considerations apply only to the perfect crystals and for
Pb$_{1-x}$Eu$_x$Te containing several percent of Eu their conclusions cannot be taken
too rigorously. In the approach  presented below we model these states by
a single, localized level of $d$ symmetry, which we refer to as the $5d$ level of
Eu. \\
\subsection{Crystal field potential in disordered crystal}
In our approach, the virtual $4f^7 \rightarrow 4f^65d^1$ transitions are
caused by the electrostatic crystal-field potential. In the present 
subsection we estimate the order of magnitude for this quantity in a disordered
Pb$_{1-x}$Eu$_x$Te crystal.

In the crystal field theory a single $4f$ or $5d$ 
electron moves in a potential that may be expanded into the following infinite series: 
\begin{equation}
\label{t4}
V_{cr}({\bf r})=\sum_{l=0}^{\infty}\sum_{m=-l}^{l}A_{lm}\left(\frac{r}{r_0}\right)^lC_{lm}(\theta,\varphi),
\end{equation}
where $r_0\approx$0.5\ \AA \  is the atomic length
unit, and the functions $C_{lm}(\theta,\varphi)$ are related to spherical
harmonics $Y_{lm}(\theta,\varphi)$ by the relation 
$C_{lm}(\theta,\varphi)=\left(\frac{4\pi}{2l+1}\right)^{1/2} 
Y_{lm}(\theta,\varphi)$. In our approach we use the simplest model of 
crystal field potential, it means we assume that the crystal field
potential is due to six point charges, each of which has charge $Ze$,
placed at $r_i\theta_i\varphi_i$, $i=1,...,6$ with respect to the
europium ion. Then the coefficients $A_{lm}$ read\cite{sugano}
\begin{equation}
\label{t5}
A_{lm}=Ze^2\sum_{i=1}^6\frac{r_0^l}{r_i^{l+1}}C_{lm}^*(\theta_i, \varphi_i),
\end{equation}
where $C_{lm}^*$ means complex conjugate to $C_{lm}$. 
In our calculations we assume that $Z=2$.

Due to the difference between Eu and Pb atoms, also the  Eu-Te
and Pb-Te distances in Pb$_{1-x}$Eu$_x$Te are different. This difference causes local lattice
deformations.  These deformations are not limited to the nearest neighborhood,
but extend over larger distances (several lattice constants). If the
concentration of Eu atoms is very small, the average distance between Eu atoms 
is large. Then, although
the crystal lattice is locally deformed, the symmetry of the Eu
surrounding is still preserved. However, with the increasing europium content,
the deformations originating from the different atoms start to overlap. Due
to a random placement of Eu atoms
in the lattice, we expect a random deviation of Eu-Te bond orientations
from those in the perfect crystal. It turns out that these deviations
cause significant ground state splittings of Eu$^{2+}$ ions.

\begin{figure}
\vspace{0.5cm}
\begin{center}
\includegraphics*[scale=0.3,angle=0]{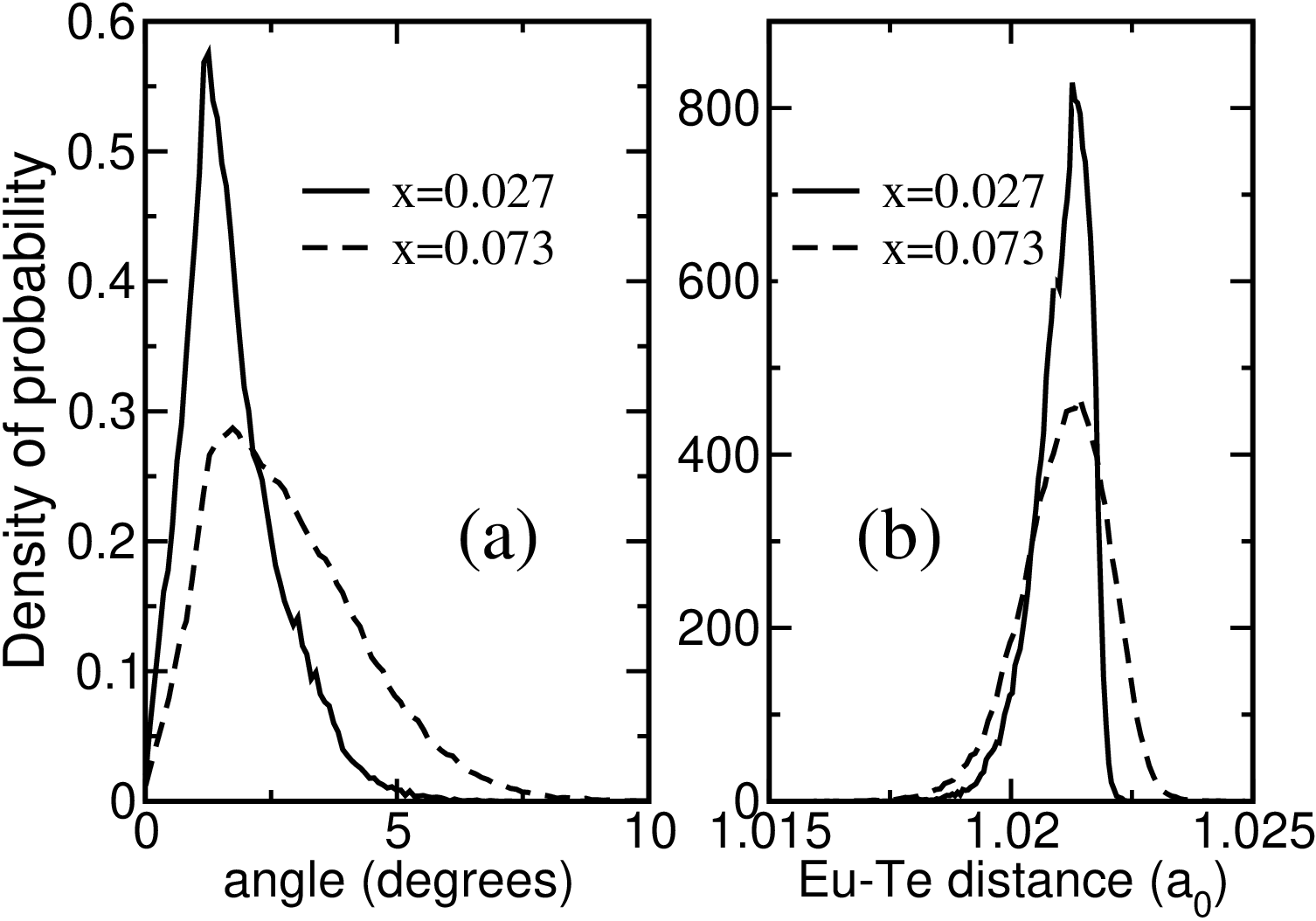}% Here is how to import EPS art
\end{center}
\caption{\label{aver1} (a) Distribution of  bonds' deviations
from the ideal crystallographic 
directions for two different Eu compositions. (b) Distribution of Eu-Te
distances in units of lattice constant of PbTe $a_0=6.46$\ \AA. }
\end{figure}

Let us consider a Pb$_{1-x}$Eu$_x$Te crystal with Eu content of several percent.
To estimate the order of magnitude of the bonds' deflections
we performed a numerical simulation. First, we modeled a perfect
lattice of crystalline PbTe containing $50^3$ unit cells of PbTe. Next, a certain
percentage of randomly chosen Pb atoms were replaced by Eu
atoms. The cations were connected to anions by springs with equilibrium
distances $d_{{\rm Eu-Te}}=3.3$~{\rm \AA } and $d_{{\rm Pb-Te}}=3.23$~{\rm
\AA}. Due to the lack of experimental data, the bond length $d_{{\rm
Eu-Te}}$ in Pb$_{1-x}$Eu$_x$Te was taken as half of EuTe lattice constant. 
After applying zero temperature Monte Carlo procedure, the
equilibrium configuration of the lattice and the
deviations of Eu-Te bonds from the perfect crystallographic
directions for every Eu atom have been found. The typical bond
deviation is of the order of several degrees. In Fig.~\ref{aver1}(a) we
plot the probability distribution for these deviations for two
different europium contents. As it may be expected, the
average deviation increases with the Eu content $x$. In
Fig.~\ref{aver1}(b) we plot the distribution for Eu-Te distances. As
we see this distribution is very well localized around the average
distance. The relative changes of the bond lengths are of the order
of 0.1\%.

The above, purely mechanical, model of disorder  serves only to
estimate the order of magnitude of deviation  due to the difference
between Eu-Te and Pb-Te bond lengths. It neglects, for
example, the difference in strength of the bonds or the angular
forces. In addition IV-VI compounds are very often disordered due to 
the presence of cation vacancies or granular structure of the
material. Due to these reasons we expect bond deviations to be larger
than estimated from Fig. \ref{aver1}.

In the description of our experimental data the degree of disorder will
serve as one of the fitting parameters. We describe now the model of
disorder which we apply to this fitting procedure.

Using a Gaussian random number generator,  we generate random deviations
of bond directions for each of the Eu atoms.  More precisely, for a
given Eu-Te bond, for example the one along (100) direction in the
perfect crystal, we generate two random angles $\beta$ and $\gamma$,
both with zero mean and the standard deviation equal to $\phi_0$. It
means that the considered bond will be characterized in spherical
coordinate system by angles $\theta=\pi/2+\beta$ and
$\phi=\gamma$. With slight modifications, a similar procedure is applied
to the remaining five bonds of the given Eu atom. In our model of
disorder we neglect changes of Eu-Te distances, because the
simulations suggest that these changes are very small. 
For the configuration of Te atoms obtained in this way 
 we calculate the crystal field potential,
Eqs.~(\ref{t4},\ \ref{t5}).

\begin{figure}
\vspace{0.5cm}
\begin{center}
\includegraphics*[scale=0.3,angle=0]{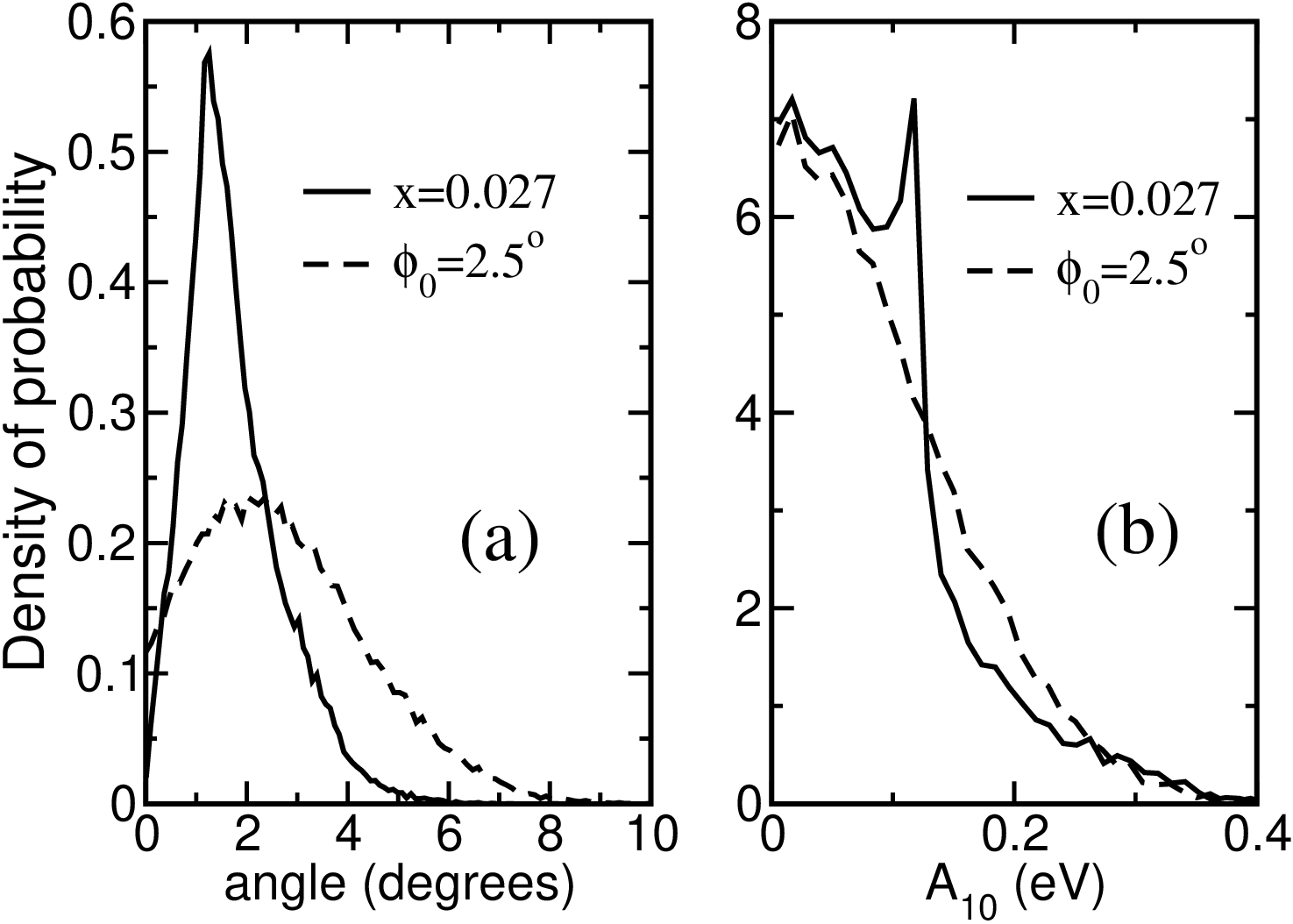}% Here is how to import EPS art
\end{center}
\caption{\label{aver2} Comparison of probability distributions for
 bond deviations (a) and  crystal field potential (b). The continuous lines
are the results of Monte Carlo simulations for a sample containing
$x$=0.027 of Eu and the broken lines are
calculated using model of disorder described in the text for $\phi_0$=2.5$^\circ$.}
\end{figure}
In Fig.~\ref{aver2} we compare the  results of Monte Carlo
procedure for a sample $x$=0.027 with those resulting from the
model of disorder introduced above for $\phi_0=2.5^{\circ}$. We see
that for such value of $\phi_0$ the differences between 
probability distributions for one of the  crystal field
coefficients, $A_{10}$, are not drastic.

From the above analysis we see that the presence of different cations
leads to a deformation of the Pb$_{1-x}$Eu$_x$Te lattice. In particular, 
deformations of nearest neighbourhoods of Eu ions cause 
splittings of Eu ions ground states. In the next subsection we discuss
physical mechanisms 
leading to the splittings.
\subsection{Mechanisms of ground state splitting of Eu$^{2+}$ ion}
In this subsection we estimate the magnitudes of the ground state
splitting due to different physical mechanisms.\cite{eps0} 
\subsubsection{Direct influence of the crystal field on 4f electrons.}
Due to the strong spin orbit coupling for $4f$ electrons, the Eu$^{2+}$ ion
is not in a pure $^8S$ state, but the higher energy states of the $4f^7$
configuration are admixed\cite{wybourne} as shown:
\begin{equation}
\label{a1}
|{\cal S}\rangle=a|^8S_{7/2}\rangle+b|^6P_{7/2}\rangle+c|^6D_{7/2}\rangle.
\end{equation}
Thus we see that the total angular momentum $L\ne 0$ and the ion may
interact with the crystal field. 
The values of the coefficients $a,b,c$ for europium may be read from
Table 5.1 of Ref.~\onlinecite{newman}: $a$=0.986, $b$=0.167, and
$c$=-0.011. For such a state, using tables of spectroscopic
coefficients,\cite{nielson} it is possible to derive matrix elements
for $\langle{\cal S}M|V_{cr}|{\cal S}M'\rangle$.\cite{liu} For
example, a term in Eq.~(\ref{t4}) proportional to $A_{20}$ 
yields the following term, $H^{20}_{MM'}$ in the effective spin Hamiltonian for
$4f^7$ shell\cite{wybourne,liu} 
\begin{equation}
\label{a2}
H^{20}_{MM'}=\delta_{MM'}\frac{2\sqrt{5}}{105}
\left(M^2-\frac{21}{4}\right)bcA_{20}\langle\left(r/r_0\right)^2\rangle,
\end{equation}
where $M$ is the projection of the total spin on a quantization axis,
$M=-7/2,...7/2$,  and the average of the second power of radius for
$4f$ shell wave function for
europium, in atomic units, is
$\langle\left(r/r_0\right)^2\rangle$=0.938 (see
Ref. \onlinecite{freeman}). 
Let us define $\Delta$ as the energy difference between the highest
and the lowest energy levels of the split ion. Then for the
Hamiltonian Eq. (\ref{a2}),
$\Delta=(24\sqrt(5)/105)bc\langle\left(r/r_0\right)^2\rangle
A_{20}$. From simulations we know that the average value of $A_{20}$ is
of the order of 0.01~eV. Thus $\Delta$ is of the order of
10$^{-2}$~meV which corresponds to 0.1~K. Other terms in the expansion,
Eq.~(\ref{t4})
give splittings of the same order of magnitude or smaller. In order to
explain the magnetic specific heat 
we need $\Delta/k_B$ to be of the order of 1 - 10~K. Thus the mechanism
based on the direct influence of the crystal field potential on $4f$ electrons
cannot explain the splitting responsible for the
experimentally observed magnetic specific heat.

\subsubsection{$4f^7\leftrightarrow$ band states hybridization}
In 1978 Barnes {\it et al.}\cite{barnes} noticed that the hybridization
between the $4f$ shell and the band states leads to a splitting of the
ground state of rare earth ions. The main idea of the model
is to consider the
excited states of the system in which the number of electrons on an ion's $4f$ shell
changes by $\pm1$. According to the Hund's rule, in the
excited states, $4f^8$ and $4f^6$, the total angular momentum of $4f$ electrons is
nonzero. Taking into account internal spin - orbit coupling, the authors of
Ref.~\onlinecite{barnes}  obtained an effective spin - lattice
interaction leading to the ground state splitting.
 Let us concentrate in this Section on processes
which lead to $4f^7 \leftrightarrow 4f^6$ transitions. These processes
may be important for Pb$_{1-x}$Eu$_x$Te because, according to the optical measurements
performed by Krenn {\it et al.}\cite{krenn}, the energy $\epsilon_1$
necessary to transfer an electron from the $4f$ shell to the conduction
band is of the order of 0.5~eV, which is rather small. The model
has been described in Ref.~\onlinecite{barnes} and  has also been 
re-derived for the octahedral symmetry of the 
ion's neighborhood as discussed in Ref.~\onlinecite{lusakowski1}. For
completeness we describe it very briefly emphasizing 
the differences necessary to account for disorder in the crystal.

The ground state of the system is a Eu ion in $4f^7$, $^8S$
configuration plus the Fermi sea of 
electrons. This eight-fold degenerate state 
of the system is characterized by $-7/2 \le M\le 7/2$, where $M$ is the
projection of $4f^7$ spin 7/2 on a quantization axis which we take
along the (001) crystallographic direction. In the excited states we
have the ion in the $4f^6$ configuration plus one additional electron
above the Fermi level, which is characterized by the set of quantum
numbers $q$. This set of quantum numbers contains the wave vector from the
first Brillouin zone, the number of the band, and the additional
quantum number necessary to fully characterize the band state. This
additional quantum number enumerates Kramers conjugated states. (For
semiconductors, for which 
the band  spin-orbit coupling may be neglected, i.e. spin of band
carrier is a good
quantum number, one may think of this additional quantum number as 
the projection of electron spin on a quantization axis. This
is not, however, the case for Pb$_{1-x}$Eu$_x$Te where the band spin-orbit coupling cannot be 
neglected.) If we assume the validity of Hund's rule for the $4f^6$ 
configuration, L=3 and S=3, the Hamiltonian for the ion in the excited
state reads:
\begin{equation}
\label{sec3_1}
H_{4f^6}=\lambda_{4f}{\bf L}\cdot{\bf S}. 
\end{equation}
The  hybridization elements are:
\begin{eqnarray}
\label{sec3_2}
& \langle L_zS_z q |H|\ M \rangle=& \\
& (-1)^{L_z+1}\sum_{\sigma=\pm\frac{1}{2}}\sqrt{\frac{7/2+2\sigma M}{7}}\delta_{S_z,M-\sigma}
\langle q|h|\phi_{-L_z\sigma}\rangle. &\nonumber
\end{eqnarray}
The state $|L_zS_z q\rangle$ is the excited state of the system in which the
projection on the quantization axis  of the total angular momentum
and spin of the ion are $L_z$ and $S_z$, respectively, and there is one
additional electron characterized by $q$ above the Fermi energy.  The
coefficient $(-1)^{L_z+1}\sqrt{\frac{7/2+2\sigma 
M}{7}}$ may be derived using
explicit forms of antisymmetric many electron functions for ion's
states $|L_zS_z \rangle$ and $|\ M \rangle$.  The
element $\langle q|h|\phi_{-L_z\sigma}\rangle$ describes hybridization
between band state $q$ and the $4f$ spin orbital
$\phi_{-L_z\sigma}$. The band wave functions are calculated within the
tight binding model\cite{bauer} and the hybridization elements between
$4f$ shell and Te $6p$ and $6s$ orbitals are described by three
constants $V_{pf\sigma}$, $V_{pf\pi}$ and $V_{sf\sigma}$. According to
Refs.~\onlinecite{harrison1} and \onlinecite{harrison2}
\begin{eqnarray}
\label{sec3_3}
V_{pf\sigma}=\eta_{pf\sigma}\frac{\hbar^2}{m_0}\frac{\left(r_pr_f^5\right)^{1/2}}{d^5}\ ,
\\
V_{pf\pi}=\eta_{pf\pi}\frac{\hbar^2}{m_0}\frac{\left(r_pr_f^5\right)^{1/2}}{d^5}\ ,
\end{eqnarray}
where $\eta_{pf\sigma}=10\sqrt{21}/\pi$,
$\eta_{pf\pi}=-15\sqrt{7/2}/\pi$, $r_p$=15.9~\AA, $r_f$=0.413~\AA \ and
$d$ is Eu-Te distance. We assume that $V_{sf\sigma}=V_{pf\sigma}$ but
the results of the calculations do not depend crucially on this
assumption. The dependence of interatomic matrix elements on the
direction of the Eu-Te bond with respect to the crystallographic axes
of the ideal crystal is calculated according the method proposed in
Ref.~\onlinecite{sharma}. The other details of calculations of the
effective spin Hamiltonian are presented in
Ref.~\onlinecite{lusakowski1}.

Due to the strong localization of the $4f$ shell, the interatomic matrix
elements responsible for the $4f$ shell  band states hybridization
are very small, of the order of 0.1~eV. This is the main reason that
there is a small splitting $\Delta$ despite the smallness of the transfer energy
$\epsilon_1$ which is approximately 0.5~eV. The average splitting
$\Delta/k_B$ is of the 
order of 0.1~K. This is again much too small a value to explain the magnetic
specific heat. 
\subsubsection{5d $\leftrightarrow$ band states hybridization}
Unlike the $4f$ orbitals of the Eu ion, the $5d$ orbitals are much more
extended in space. Thus we expect the overlapping with
neighboring Te orbitals to be much bigger. Recently one of the authors
(A. {\L}.) has proposed a new mechanism leading to the
ground state splitting of rare earth ions in which the $5d$ shell of a rare
earth ion provides a bridge between the $4f$ electrons and the rest of the
crystal.\cite{lusakowski1} More precisely, an electron from the valence band jumps
virtually onto the $5d$ level of the rare earth ion and interacts via a Heisenberg type of
exchange with the spin of the $4f$ shell. The Hamiltonian for europium in this
excited state, $4f^75d^1$ is of the following form
\begin{equation}
\label{sec3_4}
H_{4f^75d^1}=-J_{fd}{\bf S \cdot s}+\lambda_{5d} {\bf L \cdot s},
\end{equation}
where the first term describes the exchange interaction between $4f$
spin {\bf S} and the spin {\bf s} of the $5d$ electron. The second
term describes the spin orbit interaction on the $5d$ shell. 
In the more general treatment of the problem in
Ref. \onlinecite{lusakowski1} we used two spin orbit constants. Here 
we use a single spin orbit constant, 
$\lambda_{5d}$, and neglect the influence of the crystal field on $5d$
electron. For the disordered neighborhoods of europium atom which we
consider in the present paper these simplifications do not change
significantly the final results. Like the previous subsection the
influence of the surrounding comes via the hybridization  between the $5d$
and the band states. This hybridization is described by three interatomic Eu-Te matrix
elements $V_{pd\sigma}$, $V_{pd\pi}$, and $V_{sd\sigma}$ for which we
assume following Eu-Te distance dependence\cite{mg97}  
$V_{pd\sigma}=V^0_{pd\sigma}(a_0/d)^4$, $V_{pd\pi}=V^0_{pd\pi}(a_0/2d)^4$, and
$V_{sd\sigma}=V^0_{pd\sigma}(a_0/d)^{7/2}$. Here $a_0$=6.46\AA \ is
the lattice constant of PbTe and $d$ is the actual Eu-Te distance along
the given bond. The assumed values of three constants
$V^0_{pd\sigma}$=-1.5~eV, $V^0_{pd\pi}$=0.7~eV and
$V^0_{pd\sigma}$=-1.6~eV are close to the ones used in
Ref.~\onlinecite{kacman1} in calculations of EuTe band
structure.\cite{kacman} The values $J_{fd}$=0.2~eV and $\lambda_{5d}$=0.08~eV have
been taken from Table III of Ref.~\onlinecite{kasuya1}. The important
parameter of the theory is $\epsilon_2$, the energy necessary to
transfer an electron from the top of the valence band to the $5d$
level. Contrary to the case of gadolinium in PbTe where $\epsilon_2$ is of the
order of 0.4~eV, for europium we only know that it should be
larger\cite{lusakowski1}. In our calculations we have assumed
$\epsilon_2$=1~eV.

For the above values of parameters the average splitting $\Delta/k_B$ is
larger than for the two previous mechanisms, of the order of
0.5\ -\ 1~K.  It remains too small, however, to explain the magnetic specific
heat. 
\subsubsection{$4f^7\leftrightarrow 4f^65d^1$ transitions}
The last mechanism of the splitting, which we consider, is based on $4f^7\leftrightarrow
4f^65d^1$ virtual transitions. As it has been already discussed
earlier we treat $5d$ states not as the pure atomic states
but hybridized with the band states. In some sense this mechanism is
complementary to the one considered in subsection 2
because in calculations in subsection 2 the Pb$_{1-x}$Eu$_x$Te band states have not
contained $5d$ orbitals. In other words, we may say that we
add a certain amount of $5d$ to the $4f$ states.

The Hamiltonian of the model is constructed as follows. 
In the ground state of the ion there are seven electrons on the $4f$
shell. The total angular momentum is zero, the total spin equals
7/2 and this state is eight fold degenerate. As in previous sections 
$|M\rangle$ denotes a state of the ion where 
the projection of the total spin on a quantization axis in $4f^7$
configuration is $M$ ($M=-7/2,...7/2$). \\
The excited state
configuration is $4f^65d^1$. It is described by following Hamiltonian:
\begin{equation}
\label{t3}
H_{4f^65d^1}=H_{4f^6}+\lambda_{5d}{\bf l \cdot s}
- J_{fd}{\bf S \cdot s} + V_{cr} + \epsilon_0,
\end{equation}
where
\begin{equation}
\label{t3_a}
H_{4f^6}= \lambda_{4f} {\bf L \cdot S}+\lambda_{4f}^1 \left({\bf L \cdot S}\right)^2
\end{equation}
is a Hamiltonian for six electrons on $4f$ shell describing splitting of
$^7F$ state due to $4f$ spin 
orbit interaction. We assume that in the
excited configuration, six electrons on $4f$ shell behave according to
the Hund's rule, i.e. their total spin $S$=3 and the total angular momentum
$L=3$. The values of the spin orbit coupling constants
$\lambda_{4f}$=0.03~eV and $\lambda_{4f}^1$=0.0005~eV have been fitted
to describe properly a splitting of the $4f^6$ configuration calculated from first
principles (see Table VIII in Ref.~\onlinecite{visser}). 
The second term in Eq.~(\ref{t3}) describes spin orbit couplings on the $5d$
shell.   The next two terms correspond to exchange interactions
between the 
$4f$ and $5d$ spins and the crystal field
potential acting on electron on the $5d$ shell.  The last term,
$\epsilon_0$, is the energy necessary to perform the $4f^7\rightarrow
4f^65d^1$ transition. We neglect the influence
of the crystal field potential on the $4f$ electrons since we have
checked that its influence on the final result is very small.

The crystal field potential, Eq.~(\ref{t4}), enters the Hamiltonian of the
model in two places: in the term describing the excited states, Eq.~(\ref{t3}), and in
the terms describing $4f^7\leftrightarrow 4f^65d^1$
transitions.  The basis in which we describe excited states of the
ion is denoted by $|S_zL_zl_z\sigma\rangle$, where 
$S_z$ and $L_z$ correspond to projections on a quantization axis of
total spin and total angular momentum of six electrons on $4f$ shell,
respectively, while $l_z$ and $\sigma$ are $z$-th components of angular
momentum and spin of the seventh, the $5d$ electron.  Using properly
antisymmetrized wave functions or the concept of 
fractional parentage coefficients\cite{judd,nielson}, we have derived
the following form for the hybridization elements: 
\begin{eqnarray}
\label{t6}
&\langle S_zL_zl_z\sigma|\sum_{i=1}^7V_{cr}({\bf r}_i)|M\rangle= & \\
& (-1)^{L_z}\sqrt{\frac{7/2+2\sigma M}{7}}\delta_{S_z,M-\sigma}\langle 
\phi_{l_z}^{5d}|V_{cr}({\bf r})|\phi_{-L_z}^{4f}\rangle \ , & \nonumber
\end{eqnarray}
where $\phi_{l_z}^{5d}$ and $\phi_{L_z}^{4f}$ are $5d$ and $4f$
orbitals, respectively. 
The external magnetic field $B$ is
taken into account by adding to the Hamiltonian  Zeeman term:
\begin{equation}
\label{t7}
H_B=g\mu_BB\left(S_z+\sigma_z\right)+\mu_BB\left(L_z+l_z\right),
\end{equation}
where g-factor $g$=2 and $\mu_B$ is the Bohr magneton.
Calculating the matrix elements of the Hamiltonian in the basis $|M\rangle$,
$|S_zL_zl_z\sigma\rangle$  we obtain a 498$\times$498 matrix. The
eigenvalues of this matrix enable us to calculate the magnetic specific
heat according to the standard rules of statistical mechanics. The first
eight eigenvalues of the Hamiltonian matrix correspond to the split
levels of the Eu$^{2+}$ ion. The higher levels describe the $4f^65d^1$
configuration and they are separated from the lowest eight ones by an 
energy of the order of $\epsilon_0$. That is why the splitting $\Delta$ is
defined here as the difference between eighth and the lowest
eigenvalue.  The value
of the matrix element of $r$ between radial $4f$ and $5d$ wave
functions, necessary to calculate hybridization elements, may be
estimated from the Table~VI of Ref.~\onlinecite{judd1}. In
calculations we assume that $\langle 4f|r/r_0|5d\rangle$=1. The
contribution from terms in Eq.~(\ref{t4}) with $l>1$ may be
omitted since the coefficients $A_{lm}$ decay
quickly with $l$ and most important contribution comes from terms
with $l$=1. The other parameters of the model are the same as those
in the previous
sections, $J_{fd}$=0.2~eV, $\lambda_{5d}$=0.08~eV, and
$\epsilon_0$=1~eV. For such values of 
parameters of the model we obtain $\Delta/k_B$ of the order of 1-5~K.

From the analysis of different mechanisms of the splitting we conclude that
the last one, i.e. the one based on $4f^7\leftrightarrow 4f^65d^1$
transitions leads to the largest ground state splitting. This
mechanism, with the above values of parameters will be used to explain
the magnetic specific heat for a sample containing $x$=0.027 of
europium.

\subsection{Magnetic specific heat}
\begin{figure}
\vspace{0.5cm}
\begin{center}
\includegraphics*[scale=0.3,angle=0]{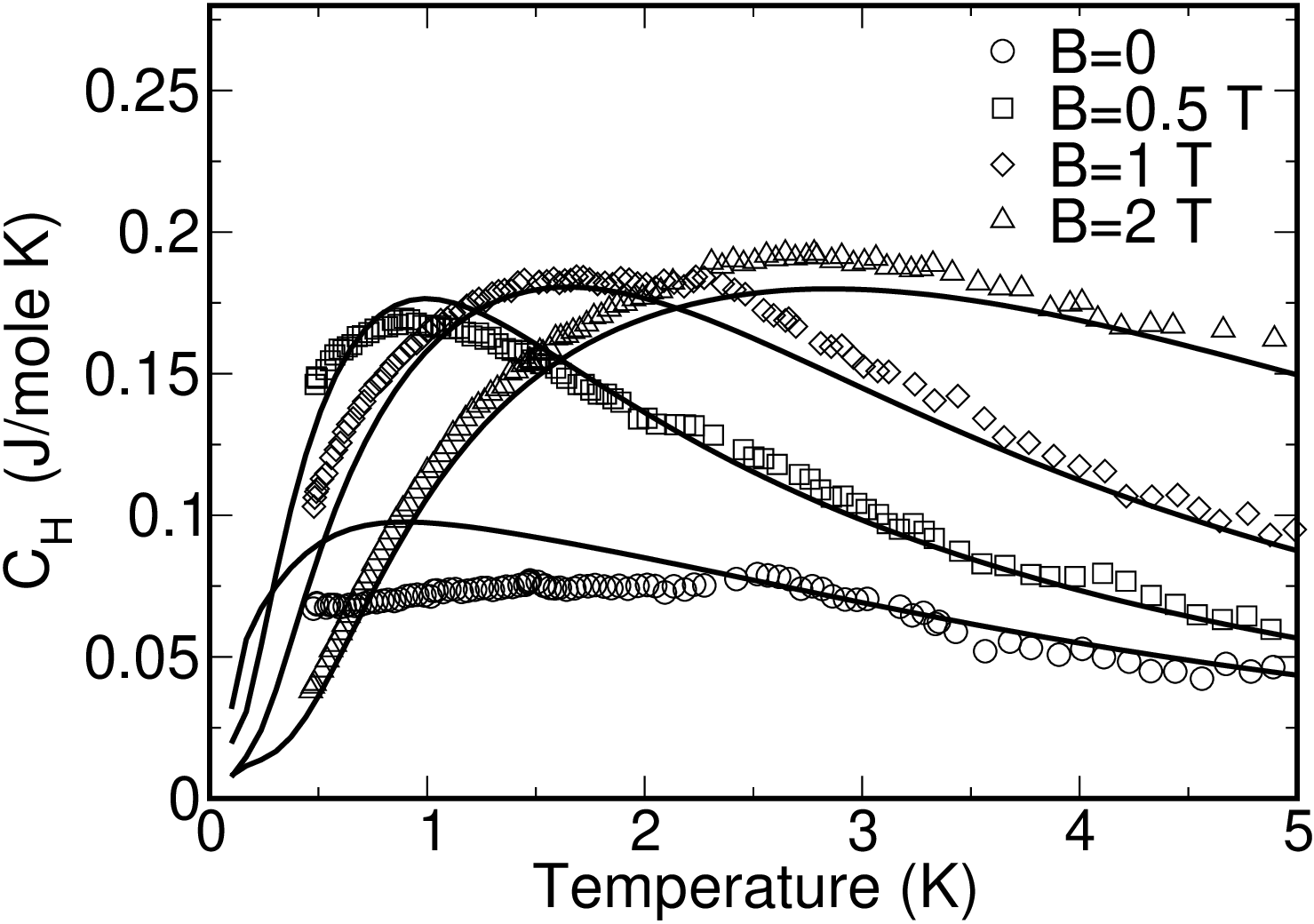}% Here is how to import EPS art
\end{center}
\caption{\label{ft4} Magnetic specific heat for sample x=0.027. Points
- experimental results. Continuous lines - theoretical calculations
for the mechanism described in Section III.B.4. }
\end{figure}

The magnetic specific heat is calculated according to standard rules
of statistical mechanics. We take into account singles, pairs and triples. We
assume that the Hamiltonian for pairs and triples is of the form given in
Eq.~(\ref{t1}) with Eu-Eu exchange integral $J/k_B$=-0.25~K.\cite{mg97,th97} For
these clusters we neglect the splitting caused by the crystal field. We do
not expect that the approximation would introduce a large error,
because a total splitting of the pairs' energy levels in zero magnetic
field 
is $28\cdot 2J$, in our case 28$\cdot2\cdot$0.25$\approx$14~K; this is the same order of
magnitude as the splitting of singles caused by the disordered crystal
field. The magnetic specific heat due to singles is calculated in the 
following way: First we generate 100 random Eu environments as  was
described in Section III~A. For each set of tellurium positions
we calculate the energy levels of the split ion and we calculate the
corresponding magnetic specific heat. As the magnetic specific heat
due to singles we take the average over these 100 samples. We found
that 100 samples are sufficient; for more samples we obtained nearly
the same result. The parameter $\phi_0$ is
treated as the fitting parameter. The best results have been obtained
for $\phi_0$=3$^{\circ}$ and they are shown in Fig.~\ref{ft4}.
\\ 

\subsection{Discussion}
In the theoretical analysis of MSH we have concentrated on the sample
containing $x$=0.027 of Eu atoms. In this sample, if one assumes a
random 
distribution of ions, more than 97\% of them are in single,
pair, and triple clusters and about 72\% are singles. As we see from
Fig.~\ref{ft4} the theoretical curves describe  the experimental
data quite well. The differences are probably 
due to the fact that the assumed model of disorder does not fully
reflect all the complexity in a real crystal. This theoretical
description has been achieved using only two fitting parameters:
$\epsilon_0$ and $\phi_0$.  All other 
parameters are known from the literature.

In the sample containing $x$=0.073 Eu, only 
about 40\% of ions are singles and more than 24\% are in clusters
containing more than
three atoms. That is why the quantitative analysis is more
difficult. However, the experimental results presented in Fig.~\ref{exp1} 
are in semiquantitative accordance with the proposed model. 

Applying Eq.~(\ref{t2}) for $B\ne 0$ and for $B=0$ we
see that the difference of the left hand sides for these two cases equals
\begin{equation}
\label{diff}
\int_{0}^{\infty}dT\frac{C_{H\ne 0}(T)}{T}-
\int_{0}^{\infty}dT\frac{C_{H=0}(T)}{T}
=k_B\ln(g_{H=0})\ ,
\end{equation}
where $g_{H=0}$ denotes the degeneracy of the ground state of the
system in zero magnetic field. (As in the previous analysis 
the ground state of the spin system in the presence of magnetic
field is nondegenerate, i.e. $\ln(g_{H\ne 0})=0$.)
Calculating the integrals using experimental data with the assumptions
discussed previously (temperature 
dependence of the specific heat for $T$<0.5~K is linear) we obtain
this difference equal to 0.17~J/mole~K. For a sample containing 0.073 of europium 40\% of
Eu ions are singles. It means that one mole of Pb$_{1-x}$Eu$_x$Te contains $0.4\cdot x \cdot
N_{A}$ of europium singles, where $N_{A}$ is the Avogadro number. Other europium ions are in larger
clusters. According to our model the ground state of each Eu single
is doubly degenerate. Thus the degeneracy of the ground
state of the spin system due to singles is $2^{0.4\cdot x \cdot 
N_{A}}$ and the right hand side of Eq. (\ref{diff}) equals $0.4\cdot
x\cdot R\ln2$.  For $x=0.073$ we obtain 0.16 \mbox{J/mole K}.   
Assuming that the ground state of Eu ions in larger
clusters is non-degenerate we obtain a very good agreement with
experimental data. However, we realize that the integrals in Eq.~(\ref{diff}) are
estimated from a limited set of data. Therefore this agreement
confirms our model only semiquantitatively.

The important problem we should consider is whether the proposed model is
consistent with the earlier magnetization and magnetic
susceptibility measurements. In the earlier theoretical description of the
experimental magnetization data, using a Brillouin-function analysis for singles, we assumed
that the Eu ions in zero magnetic field are not split.\cite{mg88,mg90,mg97}

\begin{figure}
\vspace{0.5cm}
\begin{center}
\includegraphics*[scale=0.3,angle=0]{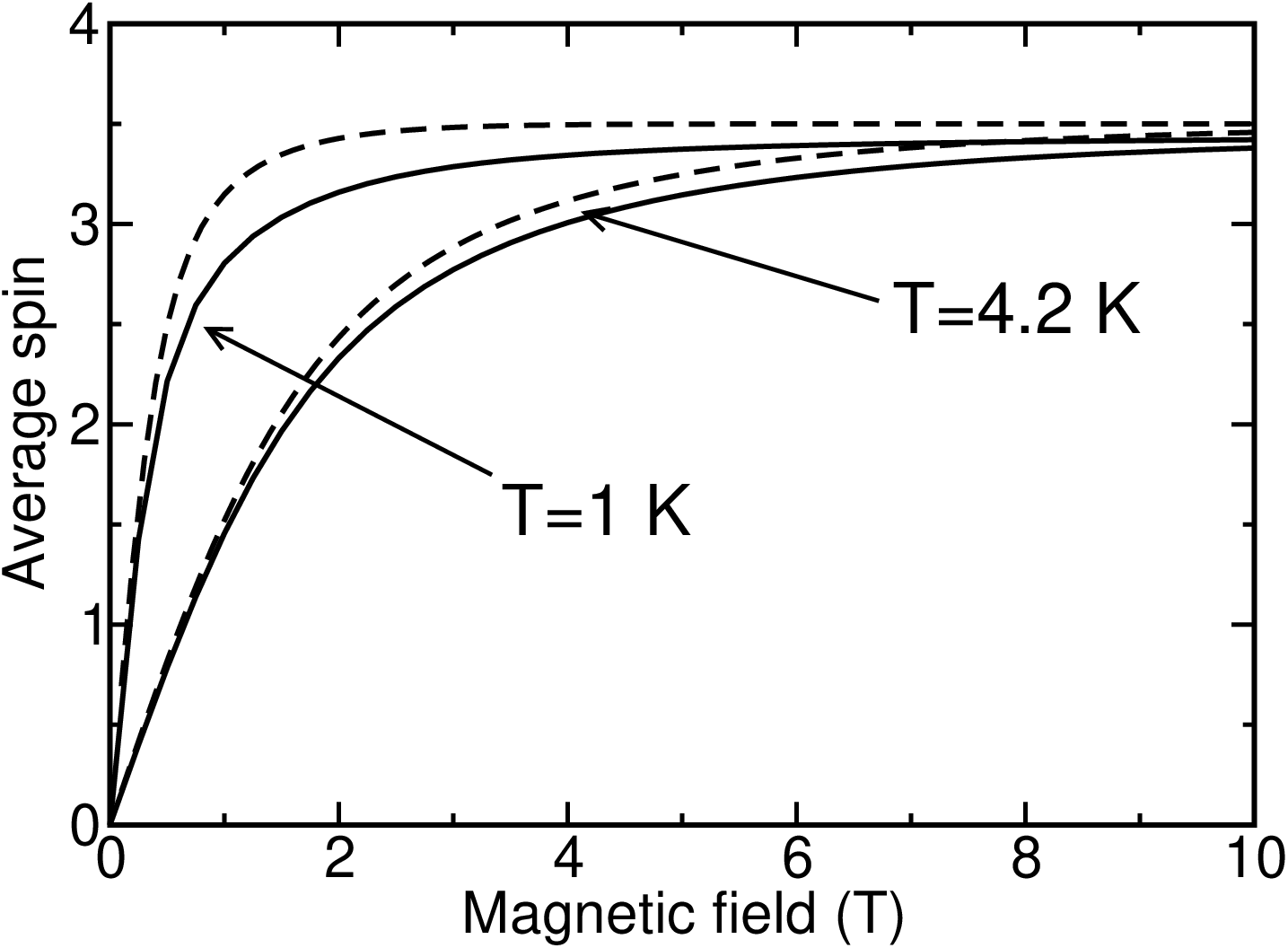}% Here is how to import EPS art
\end{center}
\caption{\label{ft5}Magnetic field dependence of average spin of a
single ion calculated according to the model - continuous lines. 
The disorder is characterized by the angle $\phi_0$=3$^{\circ}$. For
comparison, the broken lines represent plot of the Brillouin
function for spin $S$=7/2. }
\end{figure}

Fig.~\ref{ft5} shows the magnetic-field dependence of the average spin
calculated with the proposed model, continuous lines, for two
different temperatures.  The broken lines show the behavior of  Brillouin
functions calculated for spin $S$=7/2. We see that although at the
lowest temperatures the differences are
noticeable, they are small. Thus our analysis has shown that the
magnetic specific heat measurements reveal properties of the
system (density of states) which are not reflected in the
magnetization measurements. 
\section{Conclusions}
We have measured magnetic specific heat of Pb$_{1-x}$Eu$_x$Te for $x$=0.027 and
$x$=0.073. We have shown that the experimental results may be
explained assuming that the single Eu ions are split in a disordered
crystal field potential even without external magnetic
field. Analyzing the possible mechanisms of the  
splitting we have concluded that the main contribution to the
splitting comes from the virtual $4f^7\leftrightarrow 4f^65d^1$
transitions. We have also shown that the model is suitable for
description of the earlier measurements of magnetization. 
\begin{acknowledgments}
We are indebted to Prof. T. Story for helpful discussion. This work
was supported in part by the University of Maryland Center for
Superconductivity Research and by the Polish Grant PBZ-KBN-044/P03/2001.
\end{acknowledgments}

%\bibliography{theory}

\begin{thebibliography}{40}
\bibitem[{eps()}]{eps0}
\bibinfo{note}{Notice the different meanings of the symbol $\epsilon_0$ in
  subsections 2, 3 and 4. It always denotes energy of electron transfer, but
  the initial and final states in each subsection are different.}
\end{thebibliography}

\end{document}